\documentstyle[12pt]{article}
\def\ltap{\raisebox{-.55ex}{\rlap{$\sim$}} \raisebox{.4ex}{$<$}}
\def\gtap{\raisebox{-.55ex}{\rlap{$\sim$}} \raisebox{.4ex}{$>$}}
\def\gsim{\mathrel{\gtap}}
\def\lsim{\mathrel{\ltap}}

\begin{document}
\begin{titlepage}
\begin{center}
{\Large\bf Ultra-High Energy Cosmic Rays :\\ 
 a Window to Post-Inflationary
Reheating  Epoch of the Universe ? }
\vskip 1.5cm

{\large\bf V.A. Kuzmin and V.A. Rubakov}
\vskip 0.8cm
{\it Institute for Nuclear Research of Russian Academy of Sciences},\\
{\it 60th October Anniversary Prosp. 7a, Moscow 117312, Russia}\\
{\it E-mails : kuzmin@ms2.inr.ac.ru, rubakov@ms2.inr.ac.ru }
\end{center}

\begin{abstract}
We conjecture that the highest energy cosmic rays, $E > E_{GZK}$, where 
$E_{GZK} \sim 5 \cdot 10^{19}$ 
eV is the  Greisen--Zatsepin--Kuzmin cut-off 
energy of cosmic ray spectrum, may provide a unique window into 
the very early epoch of the Universe, namely, that of reheating after 
inflation, provided these cosmic rays are due to 
decays of parent superheavy long-living $X$-particles. 

These particles may  constitute a considerable fraction 
of cold dark matter  in the Universe.
We argue that the unconventionally long lifetime of the superheavy 
particles, which should be in the range of $10^{10} - 10^{22}$ years, 
might require novel particle physics mechanisms of their decays, 
such as instantons.
We propose a toy model illustrating the instanton scenario.  

Generic expected features 
of ultra-high energy extensive air showers in our scenario are 
similar to those of other top-down scenarios.
However, some properties of the upper part of the cosmic ray
spectrum make the instanton scenario 
distinguishable, at least in principle, from other ones.

\end{abstract}
\end{titlepage}

1. Among the top-down mechanisms attempting to explain the observations 
\cite{cr} of ultra-high energy (UHE) cosmic rays beyond the 
Greisen--Zatsepin--Kuzmin cut-off \cite{gzk} ,  
decays of primordial heavy particles are most obvious possibility. 
It is clearly more conventional --- at least at first sight --- 
than scenarios 
invoking topological or non-topological defects \cite{def}, though the 
latter may be a viable alternative. Heavy particles with lifetime of the 
order of the age of the Universe or greater may constitute (a substantial 
fraction of) cold dark matter, so the two observed features of the 
Universe --- ultra-high energy cosmic rays and dark matter --- may 
be related to each other.

To get an idea of the range of properties of decaying particles 
($X$-particles) that supposedly produce UHE cosmic rays, let us 
make the following simple observations. 

First, assuming sizeable hadronic component (jets) among the decay products, 
the flux of protons or gammas of energy $E$ on the Earth is estimated as
\begin{equation}
\frac{d F}{d~ln E}={\frac{1}{4\pi}}{\frac{n_{X}}{\tau_{X}}}R_{p,\gamma}
N_{j}{\frac{dN_{p,\gamma}(E)}{d~lnE}} ,  
\end{equation}     
\noindent
where $N_{j}$ is the number of jets in a typical decay, $R_{p,\gamma}$ is 
the effective distance to $X$-particles, $n_{X}$ is the number density of 
$X$-particles at the scale $R_{p,\gamma}$ , $\tau_{X}$ is the $X$-particle 
lifetime and $dN/d~lnE$ is the fragmentation function. In the following 
estimates we take $N_{j} \sim 1 - 10$ (we will soon see that large jet 
multiplicity may be favored in some decay scenarios) and 
$dN_{p,\gamma}/d~lnE \sim (\mbox {a~few})\cdot (10-100)$ 
in the energy range 
of interest, $E > (\mbox {a~few})\cdot 10^{10}$ GeV 
(the latter estimate comes 
from bold extrapolation of the fragmentation functions of ref.\cite{dok} 
to extremely high jet energies). For the effective distance we take 
$R \lsim 100$ Mpc, with understanding that the actual value of $R$ may be 
much smaller than 100 Mpc if $X$-particles are clumped. In fact, our 
conclusions will be fairly insensitive to the actual values of the 
above parameters.

The second relation is 
\begin{equation}
m_{X}<n_{X}>= \Omega_{X}\rho_{crit} ,
\label{eq:nX}
\end{equation}
\noindent
where $<n_{X}>$ is the average number density of $X$-particles and 
$\Omega_{X} \lsim 1$. If $X$-particles are clumped, the  density $n_{X}$ 
entering Eq.(\ref{eq:nX}) may be several orders of magnitude larger than 
$<n_{X}>$. Again, this uncertainty will not affect our main conclusions, 
and we set $n_{X} \sim ~<n_{X}>$ in what follows. In order to produce 
cosmic rays of energies $E \gsim (\mbox {a~few})\cdot 10^{11}$ GeV, 
the mass of 
$X$-particles is to be very large, $m_{X} \gsim 10^{13}$ GeV.

Let us now estimate the range of $X$-particle densities required. From 
Eq.(\ref{eq:nX}) we find a bound for the present $X$-particle 
density-to-entropy ratio 
\begin{equation}
n_{X}/s \lsim 10^{-21} .
\label{eq:ent}
\end{equation}
\noindent
On the other hand, to produce the observed flux of UHE cosmic rays, the 
density of $X$-particles should 
not be too small. Keeping in mind that their 
lifetime, $\tau_{X}$, cannot be much smaller than 
the age of the Universe, 
\begin{equation}
\tau \gsim 10^{10} ~yr, 
\label{eq:4+}
\end{equation}
we obtain from Eq.(\ref{eq:ent}) 
\begin{equation}
n_{X}/s \gsim 10^{-33} .
\label{eq:33}
\end{equation}
\noindent
Even though the window for $n_{X}$ is very wide, the 
estimates (\ref{eq:ent}) 
and (\ref{eq:33}) raise the issue of the production of $X$-particles 
in the early Universe. 

Alternatively, Eqs.(\ref{eq:ent}) and (\ref{eq:33}) 
may be used to place an 
upper bound on the lifetime of $X$-particles,
\begin{equation}
\tau_{X} \lsim 10^{22}~yr .
\label{eq:tau}
\end{equation}
\noindent
Again, the window for $\tau_{X}$ is wide, but the estimates (\ref{eq:33}) and 
(\ref{eq:tau}) indicate another problem, namely, that of the particle 
physics mechanism responsible for long but finite lifetime of very heavy 
particles. 

In the rest of this paper we propose possible scenarios for i) generating 
the abundance of $X$-particles in the 
range (\ref{eq:ent}), (\ref{eq:33}) 
by processes in the early Universe, and ii) explaining the lifetime of 
$X$-particles in the range (\ref{eq:4+}), (\ref{eq:tau}).

Before coming to our main points, let us stress that these two problems 
inherent in theories with very heavy and almost stable particles were 
realized long ago (see, e.g., ref.\cite{ellis} and references therein)
in different contexts. In ref.\cite{ellis} 
it was proposed that the problem i) may be solved by large entropy 
generation in the Universe after the heavy particles freeze out of 
thermal equilibrium, while their long lifetime may be due to very large 
dimension of operators responsible for the decay. 
We leave for the reader to 
judge how exotic are alternative possibilities that we discuss below.

2. If the temperature in the early Universe at some epoch exceeded the mass 
of $X$-particles and then decreased smoothly without large entropy generation, 
the freeze-out density of $X$-particles would greatly exceed the bound 
(\ref{eq:ent}). A way out of this problem is provided by inflation and 
subsequent reheating. To explain the small abundance of $X$-particles, the 
reheating temperature $T_r$ must be much smaller than $m_{X}$, so that 
$X$-particles were never at thermal equilibrium after inflation. In that case 
$X$-particles-to-entropy ratio is exponentially small, 
\begin{equation}
n_{X}/s = \mbox{const}{\cdot} \mbox{exp}(-2m_{X}/T_{r}) ,
\label{eq:6*}
\end{equation}
\noindent
where the constant depends on a number of factors (the coupling constant 
responsible for pair production of $X$-particles, the effective number of 
degrees of freedom, the ratios $m_{X}/T_{r}$ and $M_{Pl}/m_{X}$, etc.) and 
is of order $10^{-3}$ with several orders of magnitude uncertainty. As the 
dominant suppression comes from the exponential factor, the reheating 
temperature can be estimated from (\ref{eq:ent}), (\ref{eq:33}) fairly 
precisely,
\begin{equation}
T_{r}  = \left( \frac{1}{20} - \frac{1}{35}\right)m_{X} ,
\label{eq:Tr}
\end{equation}
\noindent
and should be in the range $10^{11}-10^{15}$ GeV, depending on $m_{X}$. 
Note that this range is realistic in many scenarios of inflation.

Hence, inflationary scenario can easily explain small value of the present 
density of $X$-particles in the space.
Conversely, the determination of $m_{X}$ from measurements of the upper 
end of cosmic ray spectrum would allow for rather precise estimate
of the reheating temperature. Ultra-high energy cosmic rays may indeed 
serve as a window to reheating epoch in the early Universe.

3. Explaining long lifetime of $X$-particles is much harder. 
Conventional perturbative mechanisms cannot be responsible for 
cosmologically large $\tau_{X}$ (unless very high dimension operators 
are involved \cite{ellis}), so one turns naturally to non-perturbative 
phenomena.
A well known example is instantons that produce exponentially small effects 
in weakly coupled theories. If instantons are responsible for $X$-particle 
decays, the lifetime is roughly estimated as 
\begin{equation}
\tau_{X} \sim m_{X}^{-1}\cdot \mbox{exp}(4\pi/{\alpha_{X}}) ,
\label{eq:8*}
\end{equation}
\noindent
where $\alpha_{X}$ is the coupling constant of the relevant (spontaneously 
broken) gauge symmetry. From Eqs.(\ref{eq:4+}), (\ref{eq:tau}) we find 
that the coupling constant (at the scale $m_{X}$) is 
\begin{equation}
\alpha_{X} = {\frac{1}{10}}-{\frac{1}{12}} .
\label{eq:8**}
\end{equation}
\noindent
Hence, we are lead to introduce additional non-Abelian gauge interactions 
with fairly large coupling constant at high energy scale.

To illustrate this possibility let us consider a toy model with $SU(2)_{X}$ 
gauge interactions added to the Standard Model. The $SU(2)_{X}$ gauge 
symmetry is assumed to be broken at sufficiently high energy scale. Some 
conventional quarks and leptons carry non-trivial $SU(2)_{X}$ quantum 
numbers (say, $SU(2)_{X}$ may be right-handed subgroup of a left-right 
symmetric theory, or it may be a horizontal group with generations 
forming $SU(2)_{X}$ triplets). Let there be two\footnote{The $SU(2)_{X}$ 
anomaly prevents the number of $SU(2)_{X}$ doublets from being odd.} 
left-handed $SU(2)_{X}$ fermionic doublets $X$ and $Y$ and four right-handed 
singlets, all of which are singlets under $SU(2)_{L}\times SU(3)_{c}$ of the 
Standard Model. After $SU(2)_{X}$ breaks down, 
all $X$- and $Y$-particles acquire 
large masses in a manner similar to the Standard Model. We further assume 
that $X$ and $Y$ carry different global quantum numbers, so there is no mixing 
between them. 

Under these assumptions the lightest of $X$-particles and the lightest of 
$Y$-particles (we call them $X$ and $Y$ at 
certain risk of confusing notations) 
are perturbatively stable. However, $SU(2)_{X}$ instantons induce 
effective interactions violating global quantum numbers of $X$ and $Y$. Say, 
if $X$ is heavier than $Y$, then $SU(2)_{X}$-instantons lead to the decay
\begin{equation}
X \rightarrow Y + quarks + leptons 
\label{eq:10*}
\end{equation}
\noindent
with the rate estimate given by Eq.(\ref{eq:8*}). It is this type of 
processes that may be responsible for the production of ultra-high energy 
cosmic rays.       

Let us point out a few features of this scenario which seem generic.

i) Decays induced by instantons  typically lead to multiparticle
final states. The number of quarks (jets) produced in the process
(\ref{eq:10*}) should be rather large, of order 10, 
and their distribution in energy
should be fairly flat. In principle, the spectrum of cosmic rays 
within this scenario should be distinguishable from the spectrum
predicted by other mechanisms (like two-jet decays of heavy particles 
born in the interactions of topological defects). Also, there are 
necessarily hard leptons among the decay products in the process
(\ref{eq:10*}).

ii) If $Y$-particles are indeed perturbatively stable, they are
also stable against instanton-induced interactions (because of energy
conservation and instanton selection rules). Then the dark matter in
the Universe may consist predominantly of $Y$-particles, while the
admixture of $X$-particles is small. Since the abundance of 
$Y$-particles is given by the formula similar to Eq.(\ref{eq:6*}), and 
since the density of $X$-particles is bounded  from 
above, Eq.(\ref{eq:33}), the mass splitting between $X$- and $Y$-particles
should not be large,
\begin{equation}
m_{X} < 2 m_{Y} .
\label{eq:12*}
\end{equation}
Alternatively, the Higgs sector and its interactions with fermions
may be organized in such a way that $Y$-particles are in fact
perturbatively unstable (while $X$-particles remain perturbatively
stable). In that case the heavy candidates for dark matter are
$X$-particles, and the approximate degeneracy (\ref{eq:12*}) need not 
hold.

iii) Because of the instanton selection rules, this scenario for 
the slow decay of heavy particles is rather restrictive. For example,
$X$-particles cannot be colored (otherwise the $SU(2)_{X}$ instanton
vertex would include at least three $X$-fields); if $Y$-particles 
are  stable, they cannot be colored either. This fits nicely to the
expectation that dark matter particles do not experience strong 
interactions. On the other hand, $X$-particles (and $Y$-particles, 
if stable) cannot be weak doublets for the same reason, 
so the heavy dark matter in this scenario 
does not have electroweak interactions, too. 

To conclude, an explanation of ultra-high energy cosmic ray events
beyond the GZK cut-off by decays of hypothetical heavy particles
of cosmologically long lifetime is not unrealistic from both
cosmological and particle physics points of view. Detailed study
of the upper end of  cosmic ray spectrum will provide insight into
the decay mechanism involved, and allow for the determination
of the mass of $X$-particles. As the latter has been argued to be
related to the reheating temperature, the ultra-high energy
cosmic rays may become a clue to the end-of-inflation epoch in
the early Universe.

After this work was presented at 
this Conference, we received a paper
by Berezinsky, Kachelriess and Vilenkin \cite{Berezinsky} where the
decays of heavy particles have been also considered
as the origin of cosmic rays beyond the GZK cut-off.
Their main point is that $X$-particles are expected 
to  concentrate in the galactic halo, and in this way one easily avoids the
constraints coming from the analysis 
of the cascade radiation \cite{cascade}. Their  proposal for  
long but finite lifetime of $X$-particles is that it is
due to quantum gravity (wormhole) effects.

We are indebted to V. Berezinsky, G. Farrar, W. Ochs, 
G. Pivovarov, S. Sarkar,
D. Semikoz, G. Sigl and L. Stodolsky for stimulating
discussions. The work of V.K. was supported in part by 
Russian Foundation for Basic Research grant 95-02-04911a.
The work of V.R. was supported in part by 
Russian Foundation for Basic Research grant 96-02-17449a 
and U.S. Civilian Research and Development Foundation for 
Independent States of FSU (CRDF) award RP1-187.

\end{document}